\documentclass[10pt, final, journal, twoside]{IEEEtran}

\usepackage{cite}
\usepackage{tikz}
\usepackage{scalerel}
\usetikzlibrary{svg.path}
\usepackage{mathtools,eqparbox}
\usepackage{graphicx}
\usepackage{xcolor}
\usepackage{commath}
\usepackage{amsmath}
\usepackage{amssymb}
\usepackage{amsfonts}
\usepackage{array,multirow,graphicx}
\usepackage{pgfplots}
\usepackage{subfigure}
\usepackage[utf8]{inputenc}
\newcommand{\C}{\cal{C}}
\newcommand{\rg}{R_{g_n}}
\newcommand{\Rg}{r_{g_n}}
\newcommand{\rqq}{R_{q_n}}
\newcommand{\Rqq}{r_{q_n}}
\newcommand{\ru}{R_{u}}
\newcommand{\Ru}{r_{u}}

\newcommand{\rmin}{r_{\mathrm{min}}}
\newcommand{\rearth}{r_{\mathrm{e}}}
\def\Var{{\textrm{Var}}}
\def\E{{\textrm{E}}}

\usepackage{psfrag}
\newtheorem{theorem}{Theorem}

\newtheorem{lemma}{Lemma}

\newtheorem{proposition}{Proposition}
\newtheorem{optimization problem}{Optimization Problem}
\newtheorem{corollary}{Corollary}
\pgfplotsset{compat=1.14}
\tikzstyle{arrow} = [thick,->,>=stealth]
\tikzstyle{block} = [rectangle, rounded corners, minimum width=1cm, minimum height=1cm,text centered, draw=black, fill=red!30]
\tikzstyle{input} = [circle, minimum width=2.5cm, minimum height=1cm, text centered, draw=black, fill=blue!30]
\definecolor{orcidlogocol}{HTML}{A6CE39}
\tikzset{
  orcidlogo/.pic={
    \fill[orcidlogocol] svg{M256,128c0,70.7-57.3,128-128,128C57.3,256,0,198.7,0,128C0,57.3,57.3,0,128,0C198.7,0,256,57.3,256,128z};
    \fill[white] svg{M86.3,186.2H70.9V79.1h15.4v48.4V186.2z}
                 svg{M108.9,79.1h41.6c39.6,0,57,28.3,57,53.6c0,27.5-21.5,53.6-56.8,53.6h-41.8V79.1z M124.3,172.4h24.5c34.9,0,42.9-26.5,42.9-39.7c0-21.5-13.7-39.7-43.7-39.7h-23.7V172.4z}
                 svg{M88.7,56.8c0,5.5-4.5,10.1-10.1,10.1c-5.6,0-10.1-4.6-10.1-10.1c0-5.6,4.5-10.1,10.1-10.1C84.2,46.7,88.7,51.3,88.7,56.8z};
  }
}

\newcommand\orcidicon[1]{\href{https://orcid.org/#1}{\mbox{\scalerel*{
\begin{tikzpicture}[yscale=-1,transform shape]
\pic{orcidlogo};
\end{tikzpicture}
}{|}}}}

\usepackage{hyperref} 
\newcounter{MYtempeqncnt}
\begin{document}

\title{3D Reconfigurable Intelligent Surfaces for Satellite-Terrestrial Networks}
\author{Islam~M.~Tanash and Risto~Wichman\\
Department of Information and Communications Engineering, Aalto University, Finland\\
e-mail: \{{\tt islam.tanash}, {\tt risto.wichman}\}{\tt @aalto.fi}}

\maketitle
\thispagestyle{empty} 
\begin{abstract}
This paper proposes a three-dimensional (3D) satellite-terrestrial communication network assisted with reconfigurable intelligent surfaces (RISs). Using stochastic geometry models, we present an original framework to derive tractable yet accurate closed-form expressions for coverage probability and ergodic capacity in the presence of fading. 
A homogeneous Poisson point process models the satellites on a sphere, while RISs are randomly deployed in a 3D cylindrical region. We consider nonidentical channels that correspond to different RISs and follow the $\kappa$-$\mu$ fading distribution. We verify the high accuracy of the adopted approach through Monte Carlo simulations and demonstrate the significant improvement in system performance due to using RISs. Furthermore, we comprehensively study the effect of the different system parameters on its performance using the derived analytical expressions, which enable system engineers to predict and optimize the expected downlink coverage and capacity performance analytically.
\end{abstract}


\section{INTRODUCTION}
\label{sec:intro}
\IEEEPARstart{R}{ural} areas frequently struggle to obtain high-speed internet access due to geographical barriers and a lack of infrastructure. Low Earth orbit (LEO) satellite communication has emerged as a viable solution for providing connectivity to these areas, where a line-of-sight link usually exists~\cite{Satellites-rural}, due to their ability to provide relatively low-latency and high-bandwidth connectivity compared to Geostationary satellites~\cite{Leo-survey}. However, the long transmission distances and latency between satellites and terrestrial devices may degrade information quality.  
With cutting-edge technologies like reconfigurable intelligent surfaces (RISs), the performance of LEO satellite communications can be further improved. RISs can control electromagnetic propagation, reflection, or scattering by dynamically manipulating the incident waves through a large number of small-sized passive elements to optimize communication between transmitters and receivers. LEO satellites can hence benefit from RISs to improve the system's spectral efficiency and coverage area.

Several recent studies have investigated the potential benefits of integrating RIS technology with LEO satellites, exploring a wide range of applications~\cite{pls_1,pls_2,LEO-THZ-RIS, iot,conf_ris, ICC_CON, LEO-RIS-BF,NLOS}. In particular, the authors in~\cite{pls_1,pls_2} deployed the RIS to improve the physical layer security. In~\cite{LEO-THZ-RIS}, the authors implemented the RIS in the terahertz band for LEO inter-satellite links to provide efficient inter-satellite transmission, whereas in~\cite{iot}, the authors suggested using RIS units on satellite reflectarrays to enhance IoT network connectivity with satellites networks. In~\cite{conf_ris}, the RIS was deployed in the LEO satellite on the same orbit as the source satellite to enhance the signal quality by optimizing active transmit and passive reflect beamforming. The authors in~\cite{ICC_CON} optimized user performance by mitigating interference and maximizing the weighted sum rate. A joint active and passive beamforming approach was proposed in~\cite{LEO-RIS-BF} to maximize the overall channel gain using RISs on both sides of the satellite and ground node. An RIS is used in~\cite{NLOS} to enable LEO satellite communications in non-line-of-sight channels with the aim of providing seamless cellular services.

Despite valuable contributions in the field, prior analyses are confined to small-scale systems and are missing closed-form performance measures. This can limit real-world applicability, hinder understanding, increase computational complexity, and restrict meaningful design guidelines. Motivated by the above, we leverage the demonstrated feasibility of modeling LEO satellites with stochastic geometry tools~\cite{leo_gen,Niloofar-nonhomo} to present a pioneering analytical framework distinct from conventional LEO-terrestrial satellite or RIS-assisted terrestrial networks. This framework involves characterizing the distances of both RIS's links and applying the central limit theorem (CLT) twice, setting it apart from conventional methods, while still achieving high accuracy. Furthermore, while earlier studies often focus on double-Rician channels across RIS links, our approach adopts more realistic fading conditions with the generic $\kappa$-$\mu$ fading, where various combinations of fading models can occur across the distributed RISs. We derive tractable yet extremely tight closed-form expressions for system coverage and ergodic capacity that apply to any satellite constellation, regardless of precise satellite and RIS locations.

The rest of this paper is organized as follows. Section\ref{sec:system_model} introduces a 3D RIS-assisted satellite communication system, together with the signal and fading models. The distribution of the encountered distances is presented in the same section. Section~\ref{sec: performance analysis} conducts a performance analysis in terms of the coverage probability and ergodic capacity. Section~\ref{sec:numerical} verifies the conducted analysis and further investigates the performance of the studied system. Section~\ref{sec:conc} concludes this paper.

\section{SYSTEM MODEL}
\label{sec:system_model}
 \begin{figure}[t]
    \centering
    \psfrag{2}[C][t][.9]{SAT}    
    \psfrag{3}[][][1]{$q_{n,l}$}
    \psfrag{5}[][][1]{$g_{n,l}$}
    \psfrag{c}[][][.9]{$H$}
    \psfrag{9}[][][.9]{$R_0$}
    \psfrag{7}[l][c][1]{$h_n$}
    \psfrag{6}[][][1]{$z_n$}
    \psfrag{10}[][][.9]{U}
    \psfrag{p}[][][1]{$u$}
    \psfrag{8}[c][c][.9]{RIS$_1$}
    \psfrag{b}[c][c][.9]{RIS$_2$}
    \psfrag{4}[l][t][.9]{RIS$_n$, $L_n$ REs}
    \psfrag{a}[c][c][.9]{RIS$_N$}
    \includegraphics[trim=3.5cm 6.5cm 2.5cm 14.5cm, clip=true, width=0.46\textwidth ]{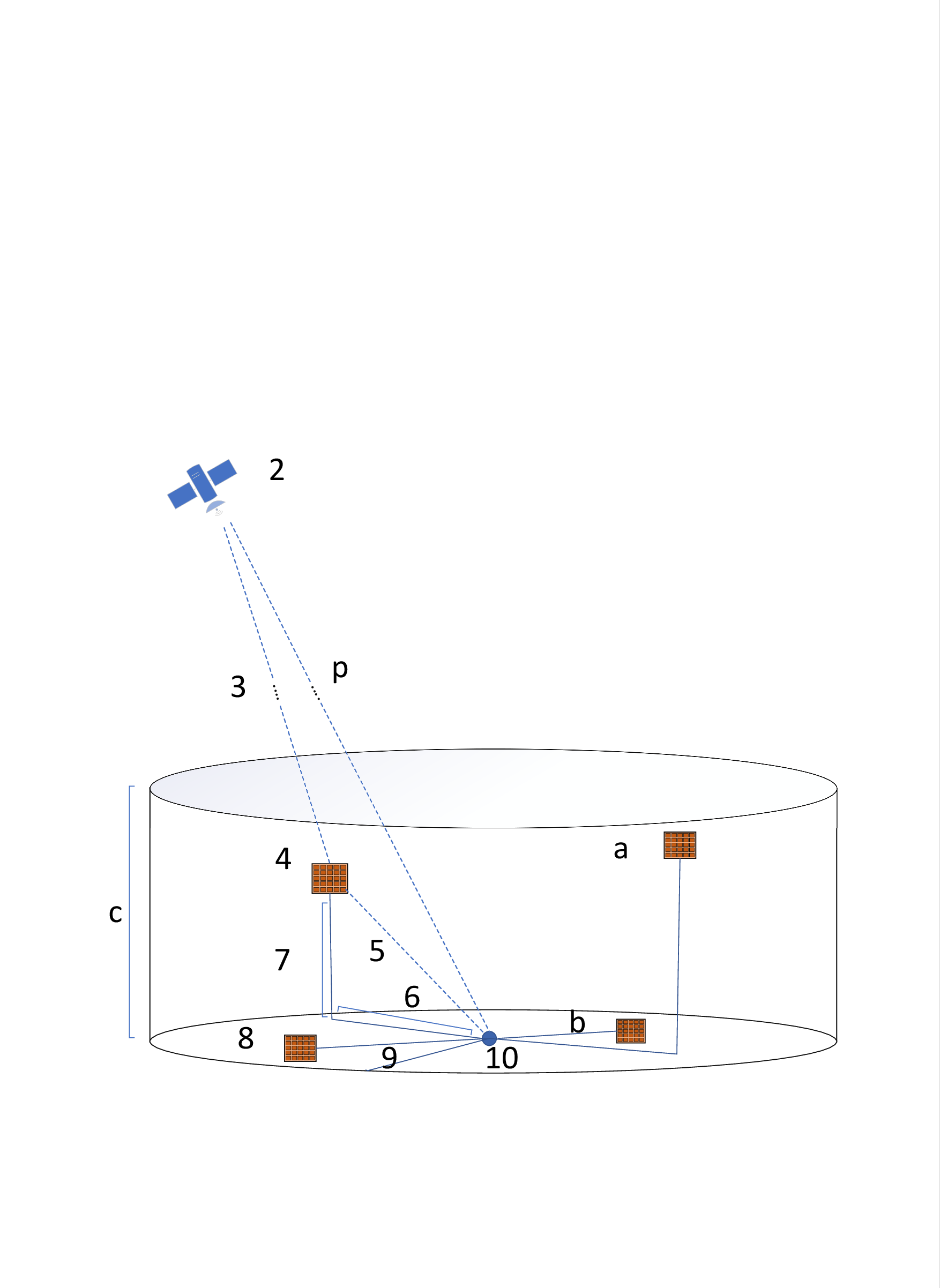}
    \textbf{\llap{\raisebox{5.9cm}{
      \includegraphics[trim=3.2cm 1.5cm 1.5cm 1.5cm, clip=true,height=1.3cm]{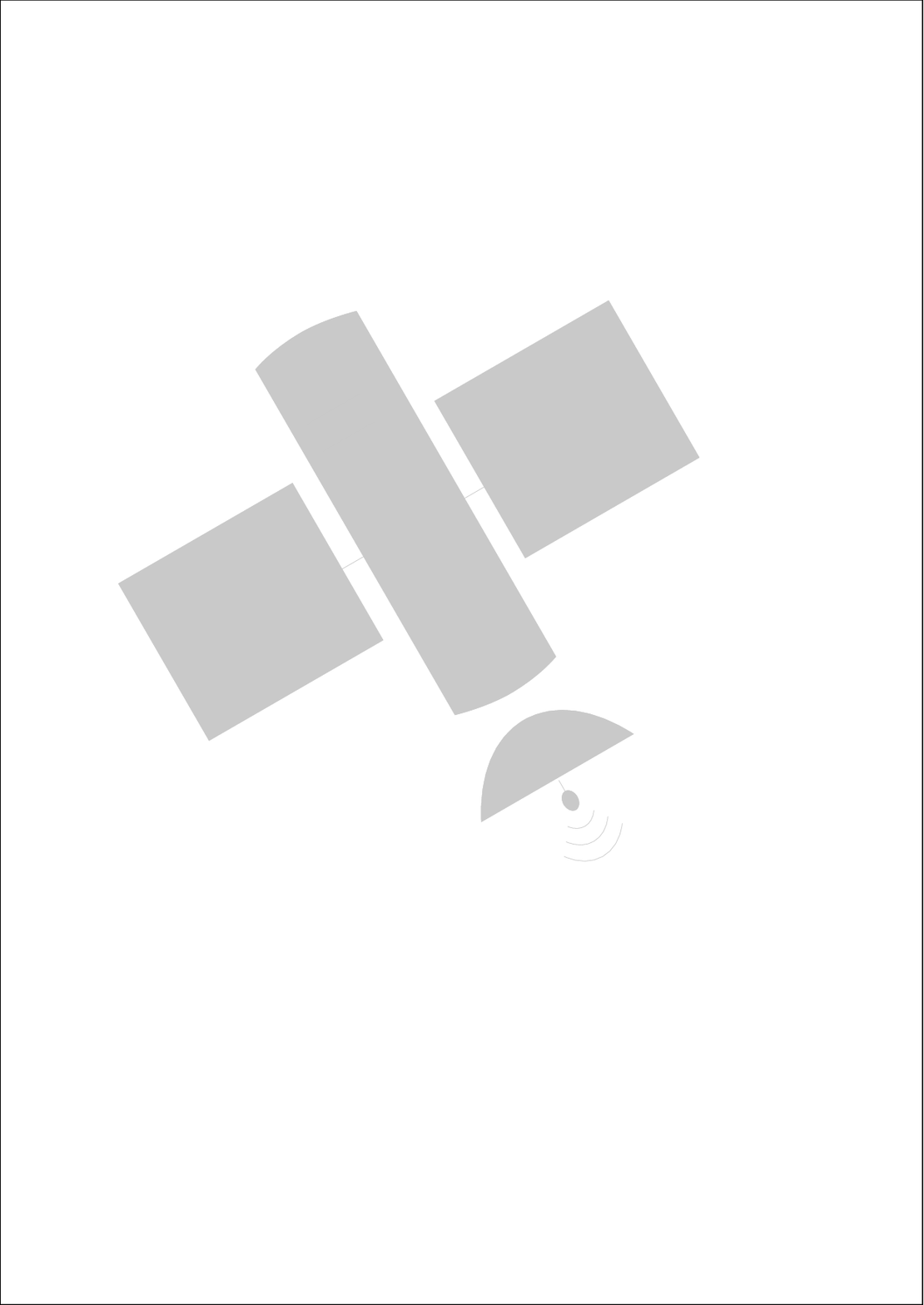}%
    }}}
        \caption{Satellite-terrestrial system with 3D RISs. Each SAT--RIS$_n$ and \mbox{RIS$_n$--U} path consists of multiple propagation paths through the $L_n$ REs.}
    \label{fig:system_model}
\end{figure}

As illustrated in Fig.~\ref{fig:system_model}, we consider an RIS-assisted satellite communication system composed of a LEO satellite constellation,
a finite 3D RISs network, wherein $N$ RISs are deployed, with the $n$th RIS (RIS$_n$) equipped with $L_n$ reflecting elements (REs) and a single-antenna user U that connects to the nearest satellite (SAT). The locations of the satellites are modeled by Poisson point process (PPP) with intensity $\lambda=\frac{M}{4\,\pi\,(\rearth + \rmin)^2}$ on a spherical surface with radius $\rearth + \rmin$, where $\rearth\approx6371$ km is the radius of Earth, $\rmin$ is the altitude of the orbit, and $M$ is the number of satellites in the constellation.   

The 3D space containing the RISs has cylindrical geometry with radius $R_0$ and maximum height $H$, inside which the RISs are distributed 
uniformly at random in both directions, i.e., vertically in $[0, H]$ and horizontally in a circle of radius $R_0$. RISs are typically constrained by height and geographical parameters, leading to a cylindrical spatial pattern. The 3D cylindrical geometry can support scalability and future expansion to accommodate additional RISs or adapt to evolving communication needs~\cite{geometry_book}. When $H=0$, RISs are distributed on the Earth's surface, simulating open rural fields. The user is located at the center of the cylindrical region's base. 
For simplicity, ideal isotropic antennas for both satellites and users are assumed. 
User U can receive signals from all distributed RISs and the direct path. The model includes the special case of an obstructed direct path between SAT and U, for which the channel coefficient $u$ below in (\ref{eq:Y}) equals zero.
\subsection{Signal Models}
The received signal at the ground user can be expressed as 
\begin{align}
\label{eq:Y}
y &= A\,s+w= \bigg(\sum_{n=1}^N A_n + \frac{u}{\ru^{\frac{\varrho}{2}}}\bigg)\,s+w,
\end{align}
with the channel response of the $n$th RIS given by
\begin{align}
\label{eq:A_n}
A_n &=\frac{\sum_{l=1}^{L_n} \,q_{n,l}\, g_{n,l}\,\zeta_{n,l}}{\rqq^{\frac{\epsilon_n}{2}}\,\rg^{\frac{\varepsilon_n}{2}}},
\end{align}
where
$s$ is the transmitted signal, $q_{n,l}$, $g_{n,l}$, and $u$ represent the fading coefficients for SAT-RIS$n$, RIS$n$-U, and SAT-U links, respectively, with $\rqq$, $\rg$, $\ru$ denoting the distances, and $\epsilon_n$, $\varepsilon_n$, $\varrho$ indicating the path-loss exponents for these links.

The additive white Gaussian noise with zero mean and variance $N_0=\E[|w|^2]$ is denoted by $w$ in (\ref{eq:Y}). Furthermore, $\zeta_{n,l}=\exp(j\theta_{n,l})$ represents the response of the $l$th RE in the $n$th RIS, with magnitude assumed to be one, and phase shift optimized to maximize the received SNR at U as $\theta_{n,l}=\angle{u}-\left(\angle{q_{n,l}}+\angle{g_{n,l}}\right)$. This paper assumes perfect channel state information (CSI). In particular, conventional CSI estimation techniques can be employed for the RISs--U links~\cite{csi-survey}, while the CSI of the RISs--SAT links can be acquired using the fact that RISs can have access to a database containing precise satellite positions, allowing them to calculate the necessary phase shifts for optimizing communication. Knowing the accurate location of satellites can help estimate the Doppler effect, after which the phase shifts of the REs can be controlled to mitigate the Doppler effect~\cite {doppler,LEO-RIS-BF}. Assuming coherent signal combining and beamforming toward the intended user, interference from other satellites via RISs is negligible. The user may, however, experience interference directly from the other satellites, which we explore in future work together with imperfect CSI.
The signal-to-noise ratio (SNR) at U is given by
\begin{align}
\label{eq:snr}
  \rho&=\rho_0\,\left|\sum_{n=1}^{N}\frac{\sum_{l=1}^{L_n} q_{n,l}\, g_{n,l}}{\rqq^{\frac{\epsilon_n}{2}}\,\rg^{\frac{\varepsilon_n}{2}}}+\frac{u}{\ru^{\frac{\varrho}{2}}} \right|^2, 
\end{align}
where $\rho_0=E_s/N_0$ is the transmit SNR with $E_s=\E[|s|^2]$. 

\subsection{ Distance distributions}
In order to assess the system's performance, it is essential to derive the distribution of encountered distances $\rqq$, $\rg$, and $\ru$, and some related statistical measures. 

\subsubsection{RIS$n$--U Link}
\begin{lemma} 
\label{lem:pdf_r_gn}
The probability density function (PDF) of the distance $\rg$ from the $n$th RIS to $U$ is given by 
\begin{equation}
\resizebox{.98\hsize}{!}{$\begin{aligned}
\label{eq:pdf_r_gn}
f_{\rg}(\Rg)
&= \begin{cases}\frac{2 \Rg^2}{R_0^2 H}, & \text {for } 0 \leq \Rg<H, \\
\frac{2 \Rg}{R_0^2}, & \text {for } H \leq \Rg<R_0, \\
\frac{2 \Rg}{R_0^2}-\frac{2 \Rg \sqrt{\Rg^2-R_0^2}}{R_0^2 H}, & \text {for } R_0 \leq \Rg \leq \psi_3,   \end{cases}
\end{aligned}$}
\end{equation}
where $\psi_3=\sqrt{R_0^2+H^2}$.
\end{lemma}
\begin{IEEEproof}
See Appendix~\ref{appen:proof_r_gn}.
\end{IEEEproof}

\begin{corollary}
When $H=0$ in Fig.~\ref{fig:system_model} (n.b., this scenario can occur in open countrysides), the cylindrical geometry will be reduced to a 2D circular surface with a PDF $f_{\rg}(\Rg)=\frac{2 \Rg}{R_0^2}$. The proof is implicitly included in Appendix~\ref{appen:proof_r_gn}. 
\end{corollary}

\begin{lemma}
\label{lem:mean_rg}
The $t$th moment of $\rg^{-\frac{\varepsilon_n}{2}}$ is given by
\begin{equation}
\resizebox{.98\hsize}{!}{$\begin{aligned}
\label{eq:mean_rg}
&\E\Big[\rg^{\frac{-t\varepsilon_n}{2}}\Big]= \frac{4 H^{2-\frac{t\varepsilon_n}{2}}}{R_0^2 (6-t\varepsilon_n)}+\frac{4 (H^{2 - \frac{t\varepsilon_n}{2}} - (H^2 + R_0^2)^{1 - \frac{t\varepsilon_n}{4}})}{R_0^2 (t\varepsilon_n-4)}\\
&-\frac{2}{3} \frac{H^2}{R_0^4} \left(H^2 + R_0^2\right)^{1 - \frac{t\varepsilon_n}{4}}\, _2F_1\left(1, \frac{5}{2} - \frac{t\varepsilon_n}{4}; \frac{5}{2}; -\frac{H^2}{R_0^2}\right),
\end{aligned}$}
\end{equation}
where $_2F_1(\cdot,\cdot;\cdot;\cdot)$ is the hypergeometric function~\cite{tableofseries}.
\end{lemma}
\begin{IEEEproof}
Substituting the piece-wise PDF derived in (\ref{eq:pdf_r_gn}) into the definition of the $t$th moment of a random variable $\rg^{-\frac{\varepsilon_n}{2}}$ as $\E\Big[\rg^{\frac{-t\varepsilon_n}{2}}\Big]=\frac{2}{R_0^2} \Big[\int_{0}^{H}\frac{ \Rg^{2-\frac{t\varepsilon_n}{2}}}{H}  \mathrm{d}\Rg + \int_{H}^{R_0} \Rg^{1-\frac{t\varepsilon_n}{2}} \mathrm{d}\Rg+ \int_{R_0}^{\psi_3} \Rg^{1-\frac{t\varepsilon_n}{2}}  \mathrm{d}\Rg- \int_{R_0}^{\psi_3} \frac{ \Rg^{1-\frac{t\varepsilon_n}{2}}\sqrt{\Rg^2-R_0^2}}{ H}  \mathrm{d}\Rg\Big]$ results in (\ref{eq:mean_rg}) after evaluating the encountered integrals.
\end{IEEEproof}
\begin{corollary}
For the special case, when $H=0$, $\E\Big[\rg^{\frac{-t\varepsilon_n}{2}}\Big]=\frac{-4 R_0^{-\frac{t \varepsilon}{2}}}{t \varepsilon-4 }$ for $ \varepsilon < \frac{4}{t}$\footnote{By integrating from $c \text{ to }R_0$, $0<c<R_0$,
when $H=0$, the $t$th moment of $R^{-\frac{\eta}{2}}$ for any $\varepsilon$ can be found as $\E\big[\rg^{\frac{-t\varepsilon_n}{2}}\big]=\frac{2}{R_0^2}\int_{c}^{R_0} \Rg^{1-\frac{t\varepsilon_n}{2}} \mathrm{d}\Rg=\frac{4 (c^{2 - \frac{t \varepsilon}{2}} - R_0^{2 - \frac{t \varepsilon}{2}})}{R_0^2 (t \varepsilon-4)}$. This implies that RISs are distributed within a disk centered on the user, with inner radius $c$ and outer radius $R_0$, rather than being deployed in a circle with radius $R_0$.}.
\end{corollary}


\subsubsection{SAT--RIS$n$ and SAT--U Links}
\begin{proposition} 
\label{prop:pdf_rq}
The distances $\rqq$ and $\ru$ have approximately the same distribution whose PDF $(f_{R}(\Rqq)$ and $f_{R}(\Ru))$ is given by
\begin{equation}
\resizebox{.98\hsize}{!}{$\begin{aligned}\label{eq:pdf_rq}
f_{R}(x)&=\frac{M x}{2 \rearth\left(\rmin+\rearth\right)} \exp \bigg(-M\bigg(\frac{x^2-\rmin^2}{4(\rmin+\rearth) \rearth}\bigg)\bigg),
\end{aligned}$}
\end{equation}
for $\rmin \leq x \leq 2 \rearth+\rmin$ and $f_{R}\left(x\right)=0$ otherwise. The shorthand $R \in \{\rqq,\ru\}$ represents both distances jointly.
\end{proposition}
\begin{IEEEproof}
Due to the substantial altitude of satellites, usually between $160$ km to $2000$ km, it is reasonable to approximate the distance between SAT and the $n$th RIS as the distance between SAT and U. Since the satellites are distributed according to homogeneous PPP, we use the PDF derived in~\cite[Eq.~14]{Niloofar-nonhomo} to state Proposition~\ref{prop:pdf_rq}.
\end{IEEEproof}

\begin{lemma}
\label{lem:mean_rq}
The $t$th moment of $R^{-\frac{\eta}{2}}$ can be found as
\begin{align}\label{eq:mean_rqq}
&\E\Big[R^{\frac{-t\eta}{2}}\Big]\nonumber\\
&=\frac{M\exp\left(\frac{M \rmin^2}{4 \rearth (\rearth + \rmin)}\right)}{4 \rearth (\rearth + \rmin)}  \bigg(\rmin^{2-\frac{t\eta}{2}}\mathrm{E}_\mathrm{\frac{t\eta}{4}}\bigg[ \frac{M \rmin^2}{4 \rearth (\rearth + \rmin)}\bigg]\nonumber\\
&- \left(2 \rearth + \rmin\right)^{2-\frac{t\eta}{2}}\mathrm{E}_\mathrm{\frac{t\eta}{4}}\bigg[\frac{M (2 \rearth + \rmin)^2}{4 \rearth (\rearth + \rmin)}\bigg]\bigg),
\end{align}
where $\eta \in \{\epsilon_n,\varrho\}$ represents the path-loss exponents of both SAT--RIS$n$ and SAT--U Links, respectively, and $\mathrm{E}_\mathrm{\frac{t\eta}{4}}$ is the exponential integral function of order $\mathrm{\frac{t\eta}{4}}$~\cite{tableofseries}. The derivation of this Lemma follows similar steps as Lemma~\ref{lem:mean_rg}.
\end{lemma}

\subsection{Fading Models}

To ensure a realistic, comprehensive, and versatile performance analysis of the proposed system model, it is crucial to consider the fact that the fading conditions experienced by both communication hops, SAT-RIS$n$ and RIS$n$-U, may differ. This variation arises because the satellite signal mostly travels in free space until it encounters the near-ground region, introducing interactions such as absorption, scattering, diffraction, and reflections that alter the characteristics of the fading channels in this region where the RISs usually reside. Moreover, the significant geographical separation of the RISs may lead to each RIS experiencing distinct fading distributions across its hops. Therefore, we adopt a generic fading model necessary for designing and optimizing communication systems for diverse and uncertain channels. In this context, we choose the normalized $\kappa$-$\mu$ distribution to model these fading scenarios.

This selection allows us to assess the system's performance over a generic distribution that encompasses Rayleigh ($\kappa=0, \mu=1$), Nakagami-$m$ ($\kappa=0, \mu=m$), Rice ($\kappa=k, \mu=1$) with $k$ being the  Rician factor, and one-sided Gaussian ($\kappa=0, \mu=0.5$), as special cases. Consequently, we can explore not only identical double fading channels for all distributed RISs but also different combinations of special cases or the generic distribution for the SAT-RIS$n$ and RIS$n$-U links across the various RISs. We also adopt the $\kappa$-$\mu$ distribution to model the direct path, covering a wide range of fading scenarios. The average gains of the normalized fading coefficients' envelopes are $\E\big[|h_{n,i}|^2\big] =\E\big[|g_{n,i}|^2\big]=\E\big[|u|^2\big]=1$. Building upon the rationale presented in~\cite{tanash-RIS}, the PDF of the end-to-end SNR in (\ref{eq:snr}) can be approximated in closed form by applying the CLT theorem twice as per the following Lemma. 
\begin{lemma}
\label{lem:pdf_snr}
The end-to-end SNR's PDF is approximated as
    \begin{align}
\label{eq:pdf_snr}
    f_{\rho}(x)&\simeq\frac{1}{2\,\beta^{\alpha}\,\Gamma(\alpha)}\,\rho_0^{-\frac{\alpha}{2}}x^{\frac{\alpha-2}{2}}\exp\left({-\sqrt{\frac{x}{\beta^2\, \rho_0}}}\right),
\end{align}
with  $\alpha=\frac{(\E[|A|])^2}{\Var[|A|]}$, $\beta=\frac{\Var[|A|]}{\E [|A|]}$, for which $A$ is the combined channel response in (\ref{eq:Y}). The mean $\E[|A|]$ and the variance $\Var[|A|]$ are given respectively in (\ref{eq:mean_A}) and (\ref{eq:var_A}) at the top of the next page.\stepcounter{equation} \stepcounter{equation}
\end{lemma}
\begin{IEEEproof}
See Appendix~\ref{appen:proof_pdf_snr}.
\end{IEEEproof}

\section{PERFORMANCE ANALYSIS}
\label{sec: performance analysis}
In this section, we derive the coverage probability and the ergodic capacity of the studied system model. 
The former defines the probability that the SNR at U exceeds the required threshold $\rho_\mathrm{th}$ for successful communication, while the latter defines the maximum rate of reliable information transmission over a time-varying communication channel.

\begin{theorem}
The probability of network coverage at U  with multiple RISs in $\kappa$-$\mu$ fading channels is given by
\begin{align}
\operatorname{P}_c&\simeq1-\frac{\gamma\left(\alpha,\sqrt{\frac{\rho_\mathrm{th}}{\rho_0\beta^2}}\right)}{\Gamma(\alpha)}.
\end{align}
\end{theorem}

\begin{IEEEproof}
$\operatorname{P}_c=\operatorname{Pr}(\rho>\rho_\mathrm{th})=1-\operatorname{Pr}(\rho\le \rho_\mathrm{th})=1-F_{\rho}(\rho_\mathrm{th})$, where $F_{\rho}(x)$ is the cumulative density function (CDF) of the end-to-end-SNR in (\ref{eq:snr}) and can be derived using the substitution $F_{\rho}(x)=F_{|A|}\Big(\sqrt{\frac{x}{\rho_0}}\Big)$. The CDF of $A$ can be found using the relation $F_{|A|}(x)=\int_{0}^{x} f_{|A|}(y)\mathrm{d} y$, where $f_{|A|}$ is given in (\ref{eq:pdf_A}). This results in $
    F_{|A|}(x)\simeq\frac{\gamma\left(\alpha,x/\beta\right)}{\Gamma(\alpha)}$,
where $\gamma(\cdot,\cdot)$ denotes the lower incomplete Gamma function.
\end{IEEEproof}

\begin{figure*}[t]
\setcounter{MYtempeqncnt}{\value{equation}}
\setcounter{equation}{8}
\begin{equation}
\resizebox{1\hsize}{!}{$\begin{aligned}
\label{eq:mean_A}
\left.\begin{matrix} 
\hspace{0 cm}\E[|A|]=\sum_{n=1}^{N}\Bigg[\overbrace{L_n\frac{\Gamma\left(\mu_{q_n}+\frac{1}{2}\right)\,\exp(-\kappa_{q_n}\,\mu_{q_n})}{\Gamma(\mu_{q_n})\,\left((1+\kappa_{q_n})\,\mu_{q_n}\right)^{\frac{1}{2}}}
 \frac{\Gamma\left(\mu_{g_n}+\frac{1}{2}\right)\,\exp(-\kappa_{g_n}\,\mu_{g_n})}{\Gamma(\mu_{g_n})\,\left((1+\kappa_{g_n})\,\mu_{g_n}\right)^{\frac{1}{2}}}\,_1F_1(\mu_{q_n}+\frac{1}{2};\mu_{q_n};\kappa_{q_n}\,\mu_{q_n})_1F_1(\mu_{g_n}+\frac{1}{2};\mu_{g_n};\kappa_{g_n}\,\mu_{g_n})}^{P_1}\nonumber\\
 \hspace{-3.25 cm}\times\Bigg(-\frac{4 H^{2-\frac{\varepsilon_n}{2}}}{R_0^2 (\varepsilon_n-6)}+\frac{4 (H^{2 - \frac{\varepsilon_n}{2}} - (H^2 + R_0^2)^{1 - \frac{\varepsilon_n}{4}})}{R_0^2 (\varepsilon_n-4)}+\frac{2}{3} \frac{H^2}{R_0^4} \left(H^2 + R_0^2\right)^{1 - \frac{\varepsilon_n}{4}}\, _2F_1\bigg(1, \frac{5}{2} - \frac{\varepsilon_n}{4}; \frac{5}{2}; -\frac{H^2}{R_0^2}\bigg)\Bigg)\nonumber\\
  \hspace{-8cm}\times\Bigg(\frac{M\exp\left(\frac{M \rmin^2}{\nu}\right)}{\nu}  \bigg(\rmin^{2-\frac{\epsilon_n}{2}}\mathrm{E}_\mathrm{\frac{\epsilon_n}{4}}\bigg[ \frac{M \rmin^2}{\nu}\bigg]- \tilde{\nu}^{2-\frac{\epsilon_n}{2}} \mathrm{E}_\mathrm{\frac{\epsilon_n}{4}}\bigg[\frac{M \tilde{ \nu}^2}{\nu}\bigg]\bigg)\Bigg)\Bigg]\end{matrix}\right\}\nonumber\text{$\E[{|A_n|}]$} \nonumber\\
 & \hspace{-560 pt}+\underbrace{\Bigg[\frac{M\exp\left(\frac{M \rmin^2}{\nu}\right)}{\nu}  \bigg(\rmin^{2-\frac{\varrho}{2}}\mathrm{E}_\mathrm{\frac{\varrho}{4}}\bigg[ \frac{M \rmin^2}{\nu}\bigg]-\tilde{\nu}^{2-\frac{\varrho}{2}} \mathrm{E}_\mathrm{\frac{\varrho}{4}}\bigg[\frac{M \tilde{\nu}^2}{\nu}\bigg]\bigg)
\frac{\Gamma\left(\mu_{u}+\frac{1}{2}\right)\,\exp(-\kappa_{u}\,\mu_{u})}{\Gamma(\mu_{u})\,\left((1+\kappa_{u})\,\mu_{u}\right)^{\frac{1}{2}}}
 \,_1F_1(\mu_{u}+\frac{1}{2};\mu_{u};\kappa_{u}\,\mu_{u})\Bigg] }_{P_2} \quad\quad\,\quad\,\quad\quad\quad\,\text{\scalebox{1.3}{ (9)}}
\end{aligned}$}
\end{equation}
\setcounter{equation}{\value{MYtempeqncnt}}
\hrulefill
\end{figure*}

\begin{figure*}[t]
\setcounter{MYtempeqncnt}{\value{equation}}
\setcounter{equation}{9}
\begin{equation}
\resizebox{1\hsize}{!}{$\begin{aligned}
\label{eq:var_A}
\Var[|A|]=&\sum_{n=1}^{N}\Bigg[\Bigg(\frac{(L_n^2-L_n)\Gamma^2\left(\mu_{q_n}+\frac{1}{2}\right)\,\exp(-2\,\kappa_{q_n}\,\mu_{q_n})}{\Gamma^2(\mu_{q_n})\,(1+\kappa_{q_n})\,\mu_{q_n}}\,\frac{\Gamma^2\left(\mu_{g_n}+\frac{1}{2}\right)\,\exp(-2\,\kappa_{g_n}\,\mu_{g_n})}{\Gamma^2(\mu_{g_n})\,(1+\kappa_{g_n})\,\mu_{g_n}}\,_1F_1^2(\mu_{h_n}+\frac{1}{2};\mu_{h_n};\kappa_{h_n}\,\mu_{h_n})\nonumber\\
&\times\,_1F_1^2(\mu_{g_n}+\frac{1}{2};\mu_{g_n};\kappa_{g_n}\,\mu_{g_n})+L_n\Biggr) \Bigg(\frac{2 H^{2-\varepsilon_n}}{R_0^2 (3-\varepsilon_n)}+\frac{2 (H^{2 - \varepsilon_n} - (H^2 + R_0^2)^{1 - \frac{\varepsilon_n}{2}})}{R_0^2 (\varepsilon_n-2)} +\frac{2H^2}{3R_0^4} \left(H^2 + R_0^2\right)^{1 - \frac{\varepsilon_n}{2}}\nonumber\\
&\times\, _2F_1\bigg(1, \frac{5-\varepsilon_n}{2}; \frac{5}{2}; -\frac{H^2}{R_0^2}\bigg)\Bigg)\Bigg(\frac{M\exp\left(\frac{M \rmin^2}{\nu}\right)}{\nu}  \Bigg(\rmin^{2-\epsilon_n}\mathrm{E}_\mathrm{\frac{\epsilon_n}{2}}\bigg[ \frac{M \rmin^2}{\nu}\bigg]-\tilde{\nu}^{2-\epsilon_n} \mathrm{E}_\mathrm{\frac{\epsilon_n}{2}}\bigg[\frac{M \tilde{\nu}^2}{\nu}\bigg]\Bigg)\Bigg)-(\E[{|A_n|}])^2\Bigg]\nonumber\\
&+\Bigg[\frac{M\exp\left(\frac{M \rmin^2}{\nu}\right)}{\nu}  \Bigg(\rmin^{2-\varrho}\mathrm{E}_\mathrm{\frac{\varrho}{2}}\bigg[ \frac{M \rmin^2}{\nu}\bigg]-\tilde{\nu}^{2-\varrho} \mathrm{E}_\mathrm{\frac{\varrho}{2}}\bigg[\frac{M \tilde{\nu}^2}{\nu}\bigg]\Bigg)-P_2^2\Bigg]. \quad\,\quad\quad\,\quad\quad\,\quad\quad\,\quad\quad\,\quad\quad\,\quad\quad\text{\scalebox{1.1}{  (10)}}
\end{aligned}$}
\end{equation}
\setcounter{equation}{\value{MYtempeqncnt}}
\hrulefill\\
$^*$note: $\nu={4 \rearth (\rearth + \rmin)}$, $\tilde{\nu}=(2 \rearth + \rmin)$, $P_2$ and $\E[{|A_n|}]$ in (\ref{eq:var_A}) are defined in (\ref{eq:mean_A}).
\end{figure*}

\begin{theorem}
The ergodic capacity of a network assisted with multiple RISs under the $\kappa$-$\mu$ fading can be written as~\cite{TCOM2}
\begin{equation}
\resizebox{1.0\hsize}{!}{$\begin{aligned}
\label{eq:capacity}
&\bar{\C}\simeq\frac{\pi}{\alpha\Gamma(\alpha)\log_e(2)}\bigg(\frac{1}{\beta^2 \rho_0}\bigg)^{{\frac{\alpha}{2}}}\csc\bigg(\frac{\pi\alpha}{2}\bigg)\, _1F_2\left(\frac{\alpha}{2};\frac{1}{2},1+\frac{\alpha}{2};-\frac{1}{4\beta^2\rho_0}\right)\\
&+\frac{\Gamma(\alpha-2)}{\Gamma(\alpha)}\frac{1}{\beta^{2}\rho_0}\, _2F_3\left(1,1;2,\frac{3}{2}-\frac{\alpha}{2},2-\frac{\alpha}{2};-\frac{1}{4\beta^2\rho_0}\right)-\frac{1}{(1 + \alpha)}\\
&\times \frac{1}{\Gamma(\alpha)}\Bigg[(2-\alpha-2\alpha^2+\alpha^3)\Gamma(\alpha-2)\Big(\log\Big(\frac{1}{\beta^2\rho_0}\Big)-2\psi^{(0)}(\alpha)\Big)\\
&+\pi\bigg(\frac{1}{\beta^2\rho_0}\bigg)^{\frac{1+\alpha}{2}}\,_1F_2\left(\frac{1}{2}+\frac{\alpha}{2};\frac{3}{2},\frac{3}{2}+\frac{\alpha}{2};-\frac{1}{4\beta^2\rho_0}\right)\sec\Big(\frac{\pi\alpha}{2}\Big)\Bigg],
 \end{aligned}$}
\end{equation}
where $\psi^{(0)}(\cdot)$ is the $0$th polygamma function and $\csc(\cdot)$ is the cosecant function.
\end{theorem}

\begin{IEEEproof}
Substituting (\ref{eq:pdf_snr}) in $\bar{\C}\triangleq\E[\log_2(1+\rho)]=\int_0^{\infty} \log_2\big(1+x\big)\,f_{\rho}(x)\,\mathrm{d}x$ and evaluating it using \cite[Eq. 15.1.3]{numericalbook} and \cite[Eq. 07.23.21.0015.01]{wolfram}, we obtain (\ref{eq:capacity}).

\end{IEEEproof}

 \section{NUMERICAL RESULTS}
 \label{sec:numerical}
 \begin{figure}[t]
\begin{center}
{\includegraphics[trim=.7cm .1cm 1cm .6cm, clip=true,width=.5\textwidth]{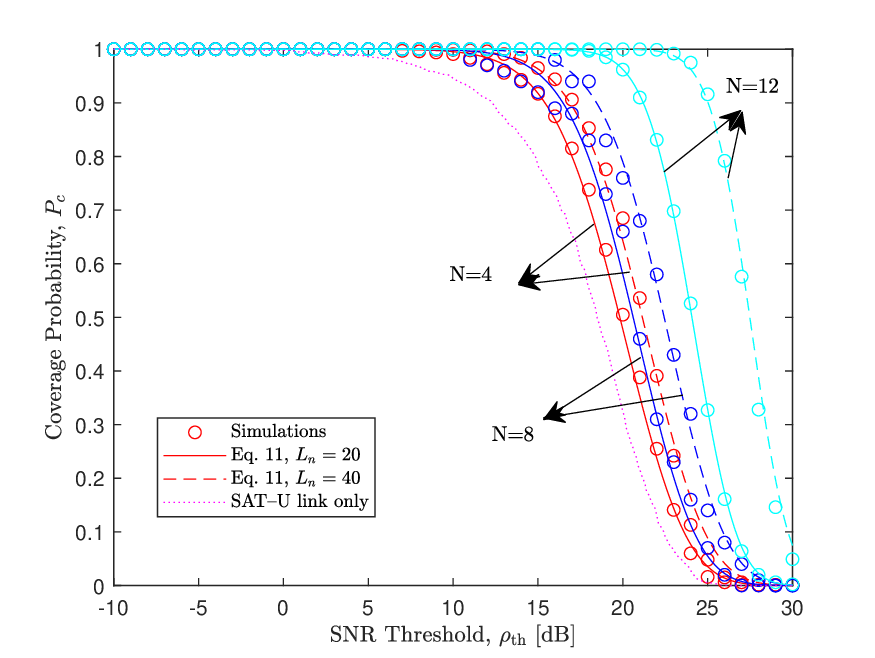}}
\caption{Effect of the number of RISs ad REs on the coverage probability.}
\label{fig:cov_N}
\end{center}
\end{figure}

\begin{figure} 
\begin{center}
{\includegraphics[trim=.6cm 0cm .1cm .5cm, clip=true,width=.5\textwidth]{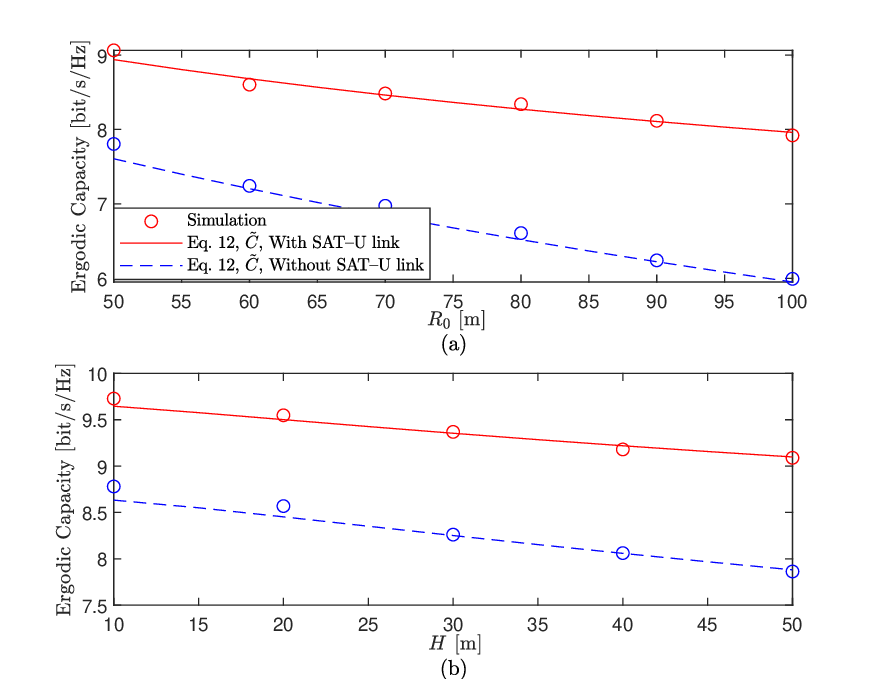}}
\caption{Impact of size of the 3D region around U on the ergodic capacity for $N=8$ and $L_n=20, n=1, 2,\ldots ,N$.}
\label{fig:cov_size}
\end{center}
\end{figure}
Based on the analytical expressions derived in Section~\ref{sec: performance analysis}, we evaluate the performance of the considered system in terms of downlink coverage probability and ergodic capacity and verify the derived expressions using Monte Carlo simulations. we set the system's parameters as $\rmin=1000$ km, $M=1000$ satellites, $E_s=10$ W, $N_0=-100$~$\mathrm{dBm}$, $\epsilon_n=\varrho=2$, $\varepsilon_n$ chosen randomly in the range $2\leq\varepsilon_n<3, n=1, 2,\ldots, N$, $\kappa_u=0, \mu_u=1$ (Rayleigh fading), $\kappa_{q_n}=1, \mu_{q_n}=2$, and $\kappa_{g_n}=3, \mu_{g_n}=3$, $n=1, 2,\ldots, N$, unless otherwise stated.

\begin{figure*}
\begin{center}
\subfigure[]{\includegraphics[trim=.7cm .1cm 1cm .7cm, clip=true,width=.49\textwidth]{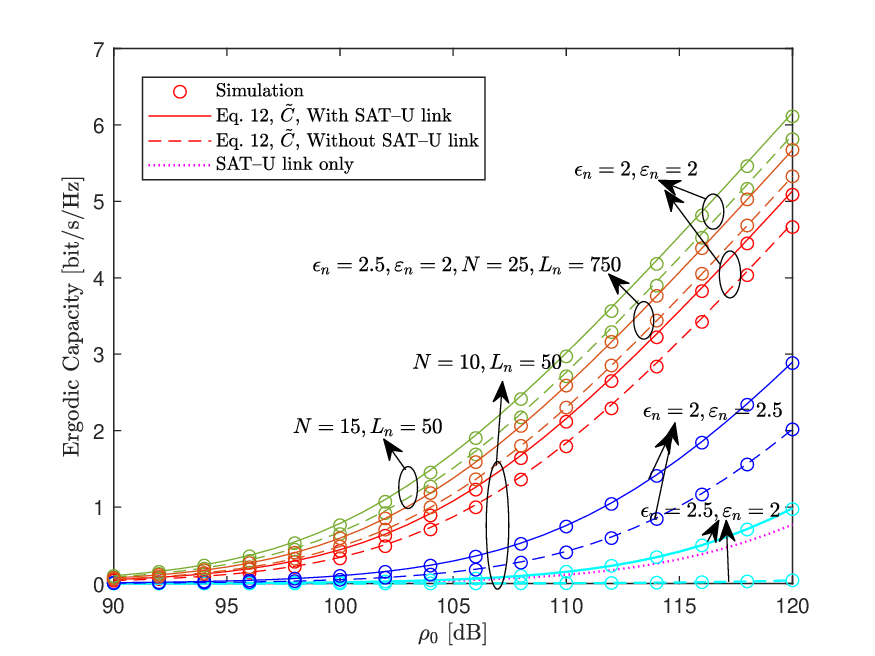}}
\subfigure[]{\includegraphics[trim=.7cm .1cm 1cm .7cm, clip=true,width=.49\textwidth]{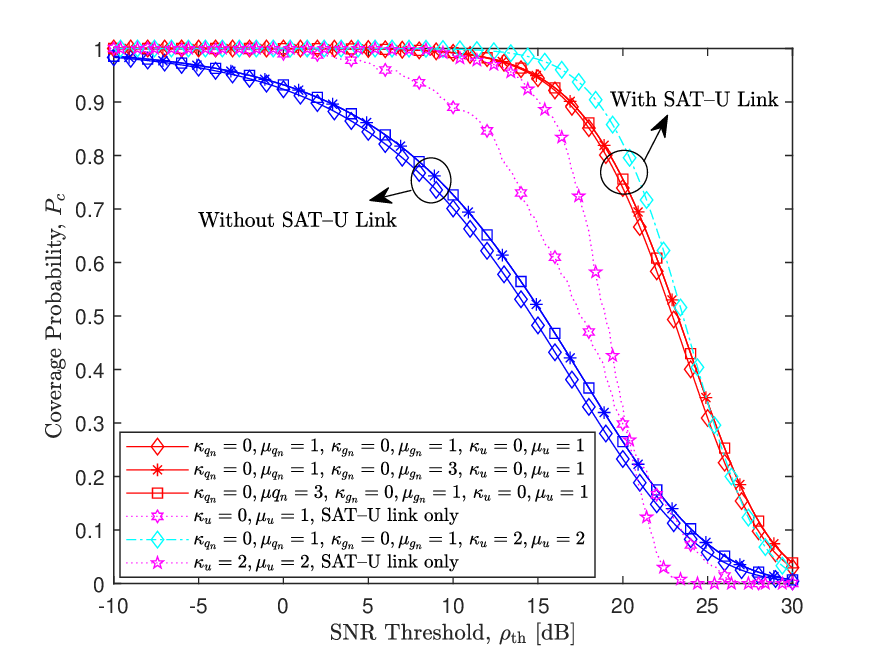}}
\caption{(a) Effect of path loss exponents on performance. (b) Effect of small-scale fading coefficients on performance for $N=8$ and $L_n=20, n=1, 2,\ldots ,N$.}
\label{fig:cap_N}
\end{center}
\end{figure*}
Figure~\ref{fig:cov_N} shows that the derived expressions and simulations are well-aligned even with a small number of RISs, e.g., $N=4$. A sufficient improvement in performance is achieved by integrating the RISs into the system compared to the case of SAT--U link only. Incorporating more REs into the RISs and/or increasing the number of RISs improves the performance even further, e.g., when increasing from $N=4$ to $N=8$, with $\rho_\mathrm{th}=20~\mathrm{dB}$ and $L_n=20$ the coverage probability is enhanced by $21.5\%$. This improvement also extends to the ergodic capacity, as shown in Fig.~\ref{fig:cap_N}(a). 

Next, in Fig.~\ref{fig:cov_size}, we study the impact of the dimensions of the cylindrical space containing the RISs on the system's behavior. Specifically, we examine the impact of increasing $R_0$ while keeping $H$ fixed and increasing $H$ while keeping $R_0$ fixed. It can be depicted from the figure that better performance is achieved by reducing both $H$ and $R_0$. This improvement is attributed to the reduced path losses when the RISs are distributed in a more confined region around the user (U).

Figure~\ref{fig:cap_N} depicts the impact of path-loss exponents and fading parameters on system performance. Generally, performance degrades with higher path-loss exponents for which the received signal suffers more severe fades. Nevertheless, SAT--RIS$_n$ links have the greatest impact on overall performance, as noted from the figure. For example, to achieve an ergodic capacity of $1$ bit/s/Hz, an increase of about $8~\mathrm{dB}$ of the transmitted power is required when $\epsilon_n=2.5, \varepsilon_n=2$, compared to when $\epsilon_n=2, \varepsilon_n=2.5$. The degradation of performance can be overcome by increasing $N$ and/or $L_n$, e.g., using $N=25$ and $L_n=750$, when $\epsilon_n=2.5, \varepsilon_n=2$ and $\rho_0=120~\mathrm{dB}$, the ergodic capacity increases by nearly $5$ bit/s/Hz, compared to when $N=10$ and $L_n=50$. In Fig.~\ref{fig:cap_N}(b), we observe that fading parameters affect both SAT--RIS$_n$ and RIS$_n$--U links similarly and as expected, improved SAT--U link fading conditions result in better coverage probability.

\section{Conclusion}
\label{sec:conc}
We proposed a LEO satellite-terrestrial system assisted with multiple RISs operating in a 3D environment and experiencing generic $\kappa$-$\mu$ fading channels, aiming to increase service quality. This work is among the first to implement tools from stochastic geometry in hybrid RISs-assisted satellite networks to provide a general mathematical framework to study such networks that are expected to be realized in the near future.
Specifically, we characterized the distance distributions of both RIS’s hops between the satellite and the user to derive analytical expressions for the downlink coverage probability and ergodic capacity. 
Our approach accommodates a wide range of popular fading models, ensuring the applicability of the derived expressions in various fading scenarios. 
Numerical results confirmed the statistical analysis and high accuracy of the performance measures. We studied the impact of 3D geometry size, fading, path-loss parameters, and the numbers of RISs and REs on system performance. Our proposed framework paves the way for precise analysis and design of more complicated future RIS-assisted satellite networks. 
\appendices
\appendices

\section{Proof of Lemma~\ref{lem:pdf_r_gn}}
\label{appen:proof_r_gn}
The 2D circular cylinder's base has uniformly distributed projections of RIS with PDF $f_{Z_n}(z_n)=\frac{2 z_n}{R_0^2}$. Their height is also uniformly distributed with PDF $f_{h_n}=\frac{1}{H}$. Thus, $\rg=\sqrt{Z_n^2+h_n^2}$ as seen in Fig.~\ref{fig:system_model}. Consequently, the CDF of $\rg$ denoted as $\Xi=F_{\rg}(\Rg)$ can be derived as
\begin{align}
\Xi&=\operatorname{Pr}\left[\rg<\Rg\right]=\operatorname{Pr}\left[Z_n<\sqrt{\Rg^2-h_n^2}\right].  \nonumber
\end{align}
This expression encounters three different cases that are illustrated in \cite[Fig. 18]{random-3D-UAV} and leads to \cite[Eq. 50]{random-3D-UAV}
\begin{equation}
\label{eq:cdf_rg}
\resizebox{1.043\hsize}{!}{$\begin{aligned}\Xi=
&\begin{cases}\int_0^{\Rg} \int_0^{\psi_1} f_{Z_n}(z_n) f_{h_n}(x) \mathrm{d} z_n \mathrm{d} x, & \text {for } 0 \leq \Rg<H, \\
\int_0^H \int_0^{\psi_1} f_{Z_n}(z_n) f_{h_n}(x) \mathrm{d} z_n \mathrm{d} x, &  \text {for } H \leq \Rg<R_0, \\
\begin{cases}\int_0^{\Rg} \int_0^{\psi_1} f_{Z_n}(z_n) f_{h_n}(x) \mathrm{d} z_n \mathrm{d} x \\
+\int_0^{\psi_2} \int_0^{R_0} f_{Z_n}(z_n) f_{h_n}(x) \mathrm{d} z_n \mathrm{d} x,\end{cases} &\text {for } R_0 \leq \Rg<\psi_3,\end{cases}
 \end{aligned}$}
\end{equation}
which results after evaluating the integrals in
\begin{equation}
\label{eq:cdf_rg}
\resizebox{1.0\hsize}{!}{$\begin{aligned}
&\Xi=
&\begin{cases}\frac{2}{3} \frac{\Rg^3}{R_0^2 H}, &  \text {for } 0 \leq \Rg<H, \\
\frac{\Rg^2}{R_0^2}-\frac{1}{3} \frac{H^2}{R_0^2}, &  \text {for } H \leq \Rg<R_0, \\
\frac{\Rg^2}{R_0^2}-\frac{1}{3} \frac{H^2}{R_0^2} -\frac{2}{3} \frac{\left(\Rg^2-R_0^2\right)^{\frac{3}{2}}}{R_0^2 H}, &\text {for } R_0 \leq \Rg<\psi_3,\end{cases}
 \end{aligned}$}
\end{equation}
where $\psi_1=\sqrt{\Rg^2-h_n^2}$, $\psi_2=\sqrt{\Rg^2-R_0^2}$, and $\psi_3=\sqrt{R_0^2+H^2}$. The PDF in (\ref{eq:pdf_r_gn}) is obtained by taking the derivative of (\ref{eq:cdf_rg}), with respect to $\Rg$.

\section{Proof of Lemma~\ref{lem:pdf_snr}}
\label{appen:proof_pdf_snr}
According to CLT, The term $\aleph_n=\left|\sum_{l=1}^{L_n} q_{n,l}\, g_{n,l}\right|$, which is a sum of double $\kappa$-$\mu$ random variables, converges to a normal distribution with mean $\E[\aleph_n]=P_1$ in (9)~\cite[Eq.~8]{tanash-RIS}, and variance~\cite[Eq.~10]{tanash-RIS}.
Next, we apply the CLT again to approximate the combined channel response $A$ in~(\ref{eq:Y}) to follow a normal distribution. The resulting PDF is further approximated using the first term of a Laguerre series expansion, i.e., the Gamma distribution~\cite{stochastic-book} to give 
\begin{align}
\label{eq:pdf_A}
    f_{|A|}(x)&\simeq\frac{x^{\alpha-1}}{\beta^{\alpha}\,\Gamma(\alpha)}\,\exp\left(-\frac{x}{\beta}\right),
\end{align}
that is substituted in $f_{\rho}(x)=\frac{1}{2 \sqrt{\rho_0 x}} f_{|A|}\left(\sqrt{\frac{x}{\rho_0}}\right)$ to yeild~(\ref{eq:pdf_snr}).
The mean in (\ref{eq:mean_A}) is computed using 
linearity and independency properties as $\E[{|A|}]=\sum_{n=1}^N E[{|A_n|}]+{\E[|u|]}{\E{[\ru^{-\frac{\varrho}{2}}]}}=\sum_{n=1}^N \E[\aleph_n]{\E[\rqq^{-\frac{\epsilon_n}{2}}]\E[\rg^{\frac{-\varepsilon_n}{2}}]}+{\E[|u|]}{\E{[\ru^{-\frac{\varrho}{2}}]}}$, for which the expectations of the distances are obtained by substituting $t=1$ in (\ref{eq:mean_rg}) and (\ref{eq:mean_rqq}). The $t$th moment of the direct path following $\kappa$-$\mu$ distribution can be found in \cite[Eq.~3]{kappa_mu_2}. 
Likewise, the variance in (\ref{eq:var_A}) is calculated as 
$\Var[|A|]=\sum_{n=1}^N\,\Var[\aleph_n\rqq^{-\frac{\epsilon_n}{2}}\rg^{\frac{-\varepsilon_n}{2}}]+\Var[\ru^{-\frac{\varrho}{2}}|u|]$. Using \cite[Eq.~1]{statistic}, together with the first ($t=1$) and second ($t=2$) monents from (\ref{eq:mean_rg}), (\ref{eq:mean_rqq}), and \cite[Eq.~3]{kappa_mu_2}, we evaluate $\Var[|A|]$ as in (\ref{eq:var_A}).

\bibliographystyle{IEEEtran}
\bibliography{Ref}

\end{document}